\documentclass[11pt,a4]{article}
\usepackage[T1]{fontenc}
\usepackage[latin1]{inputenc}
\usepackage{amsmath}
\usepackage{amssymb}
\usepackage{multirow}
\usepackage{times}
\usepackage{graphicx}
\usepackage{authblk}
\usepackage{color}

\newcommand{\captionfonts}{\scriptsize}

\makeatletter  
\long\def\@makecaption#1#2{%
  \vskip\abovecaptionskip
  \sbox\@tempboxa{{\captionfonts #1: #2}}%
  \ifdim \wd\@tempboxa >\hsize
    {\captionfonts #1: #2\par}
  \else
    \hbox to\hsize{\hfil\box\@tempboxa\hfil}%
  \fi
  \vskip\belowcaptionskip}
\makeatother   

\title{Research Letter:\\
Assessing the public health relevance of a risk factor}

\begin{document}

\author[1]{Gundula Behrens\thanks{\textit{Corresponding author:}
    Dr. Gundula Behrens, Department of Epidemiology and Preventive
    Medicine, University of Regensburg, Franz-Josef-Strauss-Allee 11, 93053 Regensburg, Germany (email: gundula.behrens@klinik.uni-regensburg.de)}}

\affil[1]{Department of Epidemiology and
    Preventive Medicine, University of Regensburg, Germany}

\maketitle
\begin{abstract}
\noindent
In a recent series of high impact public health publications, the
c-index was used as measure of prediction to assess the public health
relevance of a risk factor. I demonstrate that the c-index is an
inferior measure as compared to the classical epidemiologic measures
most commonly employed for risk prediction and public health
assessment such as disease incidence, relative risk (RR), and
population-attributable risk (PAR). I recommend using the latter
measures when assessing the public health relevance of a risk factor. 

\end{abstract}


\section*{} \label{sec:Intro}
\begin{figure}[p] 
\begin{center}
\parbox[t]{\textwidth}{\makebox[0cm]{}{
\caption{\label{fig:contour} Contour plot of the c-index in dependence
  of the risk of disease among people not exposed to a specific risk
  factor and the relative risk (RR) - or alternatively the
  population-attributable risk (PAR) - of disease associated with that
  risk factor when the prevalence of the risk factor in that
  population is (a) $50$ percent, (b) $20$ percent, or (c) $10$ percent.}}} 
\hfill 
\includegraphics[width=0.45\textwidth]{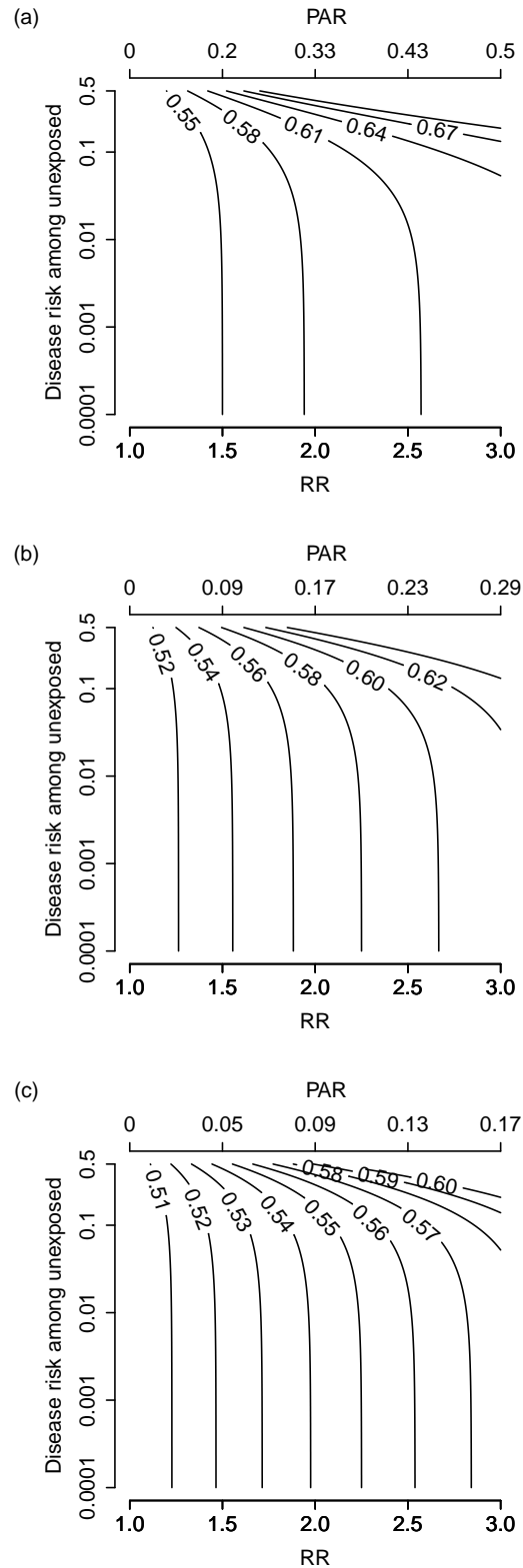} 
\end{center}
\end{figure}

\vspace{-1.5cm}
In a recent series of high impact public health publications
[1--4]\nocite{kaptogeCRP2013,diAngelantonioLipid2012,wormserSeparate2011,paynterPrediction2009},
the c-index \cite{harrellMultivariate1996} was used as measure of
prediction to assess the public health 
relevance of a risk factor. In the following, I demonstrate that
the c-index is an 
inferior measure to the classical epidemiologic measures
most commonly employed for risk prediction such as disease incidence,
relative risks (RR) and population-attributable risks (PAR). Let us
use the following notation: $p_0=$ disease incidence among those not
exhibiting the risk factor, $p_1=$ disease incidence among those
exhibiting the risk factor, $\text{RR}=p_1/p_0=$ relative risk associated with
the risk factor, $f=$ risk factor prevalence, $f_{\text{cases}}=
f\,p_1/[f\,p_1+(1-f)\,p_0]=$ risk factor prevalence among (future) cases,
$f_{\text{controls}}=f\,(1-p_1)/[ f\,(1-p_1)+(1-f)\,(1-p_0)]=$ risk factor
prevalence among (future) controls, $\text{PAR}=f\,(\text{RR}-1)/[
f\,(\text{RR}-1)+1]=$ population-attributable risk. The c-index expresses the
probability of successfully identifying the (future) case and the
(future) control among a randomly selected pair consisting of a
(future) case and a (future) control by a strategy predicting that
every subject exhibiting the risk factor is a case and every subject
not exhibiting the risk factor is a control. Because that strategy has
a probability of success of $0.5$ if both of the selected pair exhibit
or do not exhibit the risk factor, the c-index formula is given by
\begin{equation*}
\begin{split}
\text{c-index}&=0.5\times\text{probability of the selecting a case and a control
both exhibiting the risk factor}\\
&\quad+0.5\times\text{probability of the selecting a case and a control both not
exhibiting the risk factor} \\
&\quad \quad + 1\times\text{probability of the selecting a case and a
  control where the case exhibits the risk}\\
& \quad \quad \quad \quad \text{ factor and the control does not.} 
\end{split}
\end{equation*}
If we re-write that formula using $f_{\text{cases}}$ and $f_{\text{controls}}$:
\begin{equation*}
\begin{split}
\text{c-index}&=0.5\times 
f_{\text{cases}}\times f_{\text{controls}}+0.5\times (1-f_{\text{cases}})\times (1-f_{\text{controls}})+1\times
f_{\text{cases}}\times (1-f_{\text{controls}}) \\
&=0.5\times (1+f_{\text{cases}}-f_{\text{controls}}),
\end{split}
\end{equation*}
we discover a linear increase in the c-index with increasing
difference in risk factor prevalence among cases and controls. If the
risk factor prevalence in the population is fixed, then the c-index
increases with increasing RR or increasing disease incidence. Figure
\ref{fig:contour} demonstrates that, in the common epidemiologic
scenario of a disease incidence among those not exhibiting the risk
factor of less than $10$ percent and of an RR associated with the risk
factor of $1.5$, a risk factor prevalence of $20$ percent or $50$
percent would imply a substantial \text{PAR} of $9$ percent or $20$
percent, respectively, but a poor c-index not exceeding $0.55$ or
$0.56$, respectively. This demonstrates that the c-index is an
inferior measure of public health relevance to the combination of
disease incidence, relative risk, and
population-attributable risk. I therefore recommend using the latter
measures to assess the public health relevance of a risk factor. 

\bibliographystyle{unsrtabbrv} 
\bibliography{behrens_assessing_bib}

\begin{thebibliography}{1}

\bibitem{kaptogeCRP2013}
S.~Kaptoge, E.~Di~Angelantonio, L.~Pennells, A.~M. Wood, I.~R. White, P.~Gao,
  M.~Walker, A.~Thompson, N.~Sarwar, M.~Caslake, A.~S. Butterworth, P.~Amouyel,
  G.~Assmann, S.~J.~L. Bakker, E.~L.~M. Barr, E.~Barrett-Connor, E.~J.
  Benjamin, C.~Bjorkelund, H.~Brenner, E.~Brunner, R.~Clarke, J.~A. Cooper,
  P.~Cremer, M.~Cushman, G.~R. Dagenais, S.~D'Agostino, Ralph~B., R.~Dankner,
  G.~Davey-Smith, D.~Deeg, J.~M. Dekker, G.~Engstrom, A.~R. Folsom, F.~G.~R.
  Fowkes, J.~Gallacher, J.~M. Gaziano, S.~Giampaoli, R.~F. Gillum, A.~Hofman,
  B.~V. Howard, E.~Ingelsson, H.~Iso, T.~Jorgensen, S.~Kiechl, A.~Kitamura,
  Y.~Kiyohara, W.~Koenig, D.~Kromhout, L.~H. Kuller, D.~A. Lawlor, T.~W. Meade,
  A.~Nissinen, B.~G. Nordestgaard, A.~Onat, D.~B. Panagiotakos, B.~M. Psaty,
  B.~Rodriguez, A.~Rosengren, V.~Salomaa, J.~Kauhanen, J.~T. Salonen, J.~A.
  Shaffer, S.~Shea, I.~Ford, C.~D.~A. Stehouwer, T.~E. Strandberg, R.~W.
  Tipping, A.~Tosetto, S.~Wassertheil-Smoller, P.~Wennberg, R.~G. Westendorp,
  P.~H. Whincup, L.~Wilhelmsen, M.~Woodward, G.~D.~O. Lowe, N.~J. Wareham,
  K.-T. Khaw, N.~Sattar, C.~J. Packard, V.~Gudnason, P.~M. Ridker, M.~B. Pepys,
  S.~G. Thompson, and J.~Danesh.
\newblock C-reactive protein, fibrinogen, and cardiovascular risk.
\newblock {\em New England Journal of Medicine}, 367(14):1310--1320, 2013.

\bibitem{diAngelantonioLipid2012}
E.~Di~Angelantonio, P.~Gao, L.~Pennells, S.~Kaptoge, M.~Caslake, A.~Thompson,
  A.~S. Butterworth, N.~Sarwar, D.~Wormser, D.~Saleheen, C.~M. Ballantyne,
  B.~M. Psaty, J.~Sundstrom, P.~M. Ridker, D.~Nagel, R.~F. Gillum, I.~Ford,
  P.~Ducimetiere, S.~Kiechl, R.~P.~F. Dullaart, G.~Assmann, R.~B. D'Agostino,
  G.~R. Dagenais, J.~A. Cooper, D.~Kromhout, A.~Onat, R.~W. Tipping,
  A.~Gomez-de-la Camara, A.~Rosengren, S.~E. Sutherland, J.~Gallacher, F.~G.~R.
  Fowkes, E.~Casiglia, A.~Hofman, V.~Salomaa, E.~Barrett-Connor, R.~Clarke,
  E.~Brunner, J.~W. Jukema, L.~A. Simons, M.~Sandhu, N.~J. Wareham, K.-T. Khaw,
  J.~Kauhanen, J.~T. Salonen, W.~J. Howard, B.~G. Nordestgaard, A.~M. Wood,
  S.~G. Thompson, S.~M. Boekholdt, N.~Sattar, C.~Packard, V.~Gudnason, and
  J.~Danesh.
\newblock Lipid-related markers and cardiovascular disease prediction.
\newblock {\em Journal of the American Medical Association},
  307(23):2499--2506, 2012.

\bibitem{wormserSeparate2011}
D.~Wormser, S.~Kaptoge, E.~Di~Angelantonio, A.~M. Wood, L.~Pennells,
  A.~Thompson, N.~Sarwar, J.~R. Kizer, D.~A. Lawlor, B.~G. Nordestgaard,
  P.~Ridker, V.~Salomaa, J.~Stevens, M.~Woodward, N.~Sattar, R.~Collins, S.~G.
  Thompson, G.~Whitlock, and J.~Danesh.
\newblock Separate and combined associations of body-mass index and abdominal
  adiposity with cardiovascular disease: collaborative analysis of 58
  prospective studies.
\newblock {\em Lancet}, 377(9771):1085--1095, 2011.

\bibitem{paynterPrediction2009}
N.~P. Paynter, N.~R. Cook, B.~M. Everett, H.~D. Sesso, J.~E. Buring, and P.~M.
  Ridker.
\newblock Prediction of incident hypertension risk in women with currently
  normal blood pressure.
\newblock {\em American Journal of Medicine}, 122(5):464--471, 2009.

\bibitem{harrellMultivariate1996}
F.~Harrell, K.~Lee, and D.~Mark.
\newblock Multivariable prognostic models: Issues in developing models,
  evaluating assumptions and adequacy, and measuring and reducing errors.
\newblock {\em Statistics in Medicine}, 15(4):361--387, 1996.

\end{thebibliography}

\end{document}